\documentclass{INTERSPEECH2023}

\usepackage{cite}
\usepackage{setspace}
\usepackage{bm}
\usepackage{multirow}
\usepackage{subfig}
\usepackage{comment}
\usepackage{amssymb}
\usepackage{hyperref}
\usepackage{colortbl}
\usepackage{xcolor}
\usepackage{setspace}
\usepackage{arydshln}
\usepackage{threeparttable}

\usepackage{algorithm}
\usepackage{algpseudocode}

\algnewcommand\algorithmicforeach{\textbf{for each}}
\algdef{S}[FOR]{ForEach}[1]{\algorithmicforeach\ #1\ \algorithmicdo}
\algnewcommand\algorithmicswitch{\textbf{switch}}
\algnewcommand\algorithmiccase{\textbf{case}}
\algnewcommand\algorithmicassert{\texttt{assert}}
\algnewcommand\Assert[1]{\State \algorithmicassert(#1)}%
\algdef{SE}[SWITCH]{Switch}{EndSwitch}[1]{\algorithmicswitch\ #1\ \algorithmicdo}{\algorithmicend\ \algorithmicswitch}%
\algdef{SE}[CASE]{Case}{EndCase}[1]{\algorithmiccase\ #1}{\algorithmicend\ \algorithmiccase}%
\algtext*{EndSwitch}%
\algtext*{EndCase}%

\newcommand{\nm}[1]{{\color[HTML]{000000}{#1}}}

\interspeechcameraready 

\title{Diffiner: A Versatile Diffusion-based Generative Refiner\\for Speech Enhancement}
\name{
\begin{tabular}{@{}c@{}}
Ryosuke Sawata$^\dagger$ \qquad 
Naoki Murata$^\ddagger$ \qquad
Yuhta Takida$^\ddagger$ \qquad
Toshimitsu Uesaka$^\ddagger$ \\
Takashi Shibuya$^\ddagger$ \qquad 
Shusuke Takahashi$^\dagger$ \qquad
Yuki Mitsufuji$^\ddagger$
\end{tabular}}
\address{$^\dagger$Sony Group Corporation, Tokyo, Japan\\
$^\ddagger$Sony Research, Tokyo, Japan}
\email{\{Ryosuke.Sawata, Naoki.Murata, Yuta.Takida, Toshimitsu.Uesaka, Takashi.Tak.Shibuya, Shusuke.Takahashi, Yuhki.Mitsufuji\}@sony.com}

\begin{document}
\maketitle

\begin{abstract}
Although deep neural network (DNN)-based speech enhancement (SE) methods outperform the previous non-DNN-based ones, they often degrade the perceptual quality of generated outputs.
To tackle this problem, we introduce a DNN-based generative refiner, Diffiner, aiming to improve perceptual speech quality pre-processed by an SE method.
We train a diffusion-based generative model by utilizing a dataset consisting of clean speech only.
Then, our refiner effectively mixes clean parts newly generated via denoising diffusion restoration into the degraded and distorted parts caused by a preceding SE method, resulting in refined speech.
Once our refiner is trained on a set of clean speech, it can be applied to various SE methods without additional training specialized for each SE module.
Therefore, our refiner can be a versatile post-processing module w.r.t. SE methods and has high potential in terms of modularity.
Experimental results show that our method improved perceptual speech quality regardless of the preceding SE methods used.
Our code is available at \url{https://github.com/sony/diffiner}.
\end{abstract}

\section{Introduction}
\label{sec:intro}
In the field of speech enhancement (SE), DNN-based methods have drastically improved the performance of conventional ones in terms of signal-to-noise ratio (SNR) \cite{conv_tasnet, rnn_se_1, crnn_hybrid_se_3}.
However, in some cases they tend to degrade qualities of speech such as naturalness and perceptual quality for human listening \cite{sp_perceptualquality}.
Because the inputs for downstream applications (e.g., ASR and telecommunication system) should ideally be clean and high-quality speech, it has been reported that the speech processed by the aforementioned DNN-based methods often degrade their performances \cite{DNNbased_SingleSE_degradeASR_2, DNNbased_SingleSE_degradeASR_3, DNNbased_SingleSE_degradeASR_5}.

There are two main approaches to solving this problem: a) a DNN learning strategy that aims to improve both SNR and perceptual speech quality, and b) optimizing a preceding SE model in terms of the downstream application's criterion.
Regarding approach a), there are some studies that introduce criteria related to perceptual quality of human into loss function in order for the target DNN to deal with it \cite{percep_loss_se_1, percep_loss_se_2, metricgan}.
For example, Fu \textit{et al.} proposed a DNN-based SE method that can be optimized by utilizing an arbitrary metric related to the perceptual quality for human listening \cite{metricgan}, e.g., perceptual evaluation of speech quality (PESQ) \cite{pesq} and short-time objective intelligibility (STOI) \cite{stoi}, by using a framework of generative adversarial network (GAN).
However, although PESQ and STOI are correlated with perceptual speech quality, optimizing the target SE model on the basis of these metrics does not always improve the actual quality perceived by humans because the mechanisms of PESQ and STOI are not perfectly equal to human listening \cite{dnnse_percep_human}.
Shi \textit{et al.} and Liu \textit{et al}. hypothesized that synthesizing conditioned speech would improve perceptual quality and proposed using a vocoder for the SE task to generate clean speech \cite{se_vocoder_1, se_vocoder_2}.
However, training the vocoder with noisy speech tends to be more laborious than in the case of the SE model only, and it often degrades the final perceptual quality.
Meanwhile, in terms of approach b), some studies have attempted to optimize SE models so that a criterion of the following application is maximized.
For instance, the joint training connects the DNNs of the SE and ASR models and trains the connected model as one DNN in terms of the ASR's criteria \cite{joint_se_am_1, joint_se_am_2}.
However, because this approach requires training the SE model for each following model, the learning requires much effort.
Therefore, a new scheme is desired that can improve the perceptual speech quality without the laborious data collection and training.

To remove the distortions from SE outputs, we focus on the utilization of deep generative models built on only clean speech.
Although there are some SE methods using deep generative model built on both noisy and clean speeches \cite{segan, welker2022speech, richter2022speech}, which they are same as the traditional DNN-based SE methods, a generative model should be built on the target domain, i.e., clean speeches in our case, and thus using noisy speeches may degrade its performance.
To be more specific, we consider the task of SE as a generative task, where generative models are expected to detect degraded parts and refine them by mixing generated clean parts effectively.
Recently, many types of generative models such as the GAN~\cite{goodfellow2014generative}, variational auto-encoder (VAE)~\cite{kingma2013auto}, flow-based models~\cite{rezende2015variational}, and denoising diffusion-based models~\cite{ho2020denoising} have been proposed.
More recently, denoising diffusion-based generative models in particular have been extensively studied~\cite{ho2020denoising, song2020denoising, dhariwal2021diffusion}. 
Kawar \textit{et al.} devised Denoising Diffusion Restoration Models (DDRM) as an effective way to use the pre-trained denoising diffusion-based generative model for general linear inverse problems~\cite{kawar2022denoising}. DDRM can be applied to various tasks (e.g., image super-resolution, inpainting, and colorization) without any specialized retraining for each task.

Inspired by the successful adaptation of diffusion-based restoration models, we propose a diffusion-based generative refiner, Diffiner, for pre-processed speech.
Specifically, we first train a diffusion-based generative model by utilizing a dataset consisting of clean speech only.
Then, we effectively mix clean parts newly generated via the above DDRM-based framework into degraded and distorted ones caused by a preceding SE method, resulting in refinement.
More specifically, our method calculates weights based on the output of a preceding SE model and does the above mixing by utilizing the weights.
Consequently, Diffiner can boost the perceptual speech quality without no new additional noise datasets since Diffiner can be trained by using only clean speech.
It is worth noting that even the same clean speech used to train the preceding SE model can be used for training Diffiner, i.e., without any collection of new speech and noise datasets.
Furthermore, once our model is trained only on a set of clean speech, it can be applied to various SE methods without additional training specialized for each SE module.
In summary, even though our model is built on only clean speech and does not require retraining the preceding SE model, the perceptual speech quality can be refined.
Therefore, our method is versatile w.r.t. SE methods and has high potential in terms of modularity.
In our experiments, we show that our method effectively improves speech quality regardless of the SE method used for pre-processing.

\section{Preliminaries}
\label{sec:2}
In this section, we revisit diffusion-based generative models and a related diffusion-based method for general linear inverse problems.

\subsection{Diffusion-based generative models}
\label{sec:DDPM}
Denoising diffusion probabilistic models~\cite{ho2020denoising, nichol2021improved}, which we refer to as diffusion-based generative models in this paper, are latent variable models with the latents $\bm{x}^{(1)}, \dots, \bm{x}^{(T)}$, where $T$ is the number of time steps. The latents have the same dimensionality as the data $\bm{x}^{(0)} \sim p_{\text{data}}(\bm{x}^{(0)})$. Their joint distribution is defined as a Markov chain, and the transitions starting at $p(\bm{x}^{(T)})$, which is a Gaussian distribution, follow the Gaussian transition, as
\begin{align}
    p_{\theta}(\bm{x}^{(0:T)}) &= p(\bm{x}^{(T)})\prod_{t=1}^{T}p_{\theta}(\bm{x}^{(t-1)}|\bm{x}^{(t)}), \\ 
    p_{\theta}(\bm{x}^{(t-1)}|\bm{x}^{(t)})&=\mathcal{N}(\bm{x}^{(t-1)}; \bm{\mu}_{\theta}^{(t)}(\bm{x}^{(t)}), \bm{\Sigma}_{\theta}^{(t)}(\bm{x}^{(t)})).
\end{align}
The parameter of the model $\theta$ is learned to generate the data distribution following such a Markov chain so that it is consistent with the following inference distribution $q$: 
\begin{align}
    q(\bm{x}^{(1:T)}|\bm{x}^{(0)}) &= \prod_{t=1}^{T}q(\bm{x}^{(t)}|\bm{x}^{(t-1)}), \nonumber \\ 
    q(\bm{x}^{(t)}|\bm{x}^{(t-1)}) &= \mathcal{N}(\bm{x}^{(t)}; \sqrt{1-\beta_{t}}\bm{x}^{(t-1)}, \beta_{t}\bm{I}).
\end{align}
The former process is called the \textit{reverse process}, whereas the latter is called the \textit{forward process}. The frequently used parameterization of the mean $\bm{\mu}_{\theta}$ is \begin{align}
    \bm{\mu}_{\theta}^{(t)}(\bm{x}^{(t)}) = \frac{1}{\sqrt{\alpha_{t}}}\left(\bm{x}^{(t)}-\frac{\beta_{t}}{\sqrt{1-\overline{\alpha}_{t}}}\bm{\epsilon}_{\theta}^{(t)}(\bm{x}^{(t)})\right),
    \label{eq:def_mu}
\end{align}
where $\alpha_{t}:=1-\beta_{t}$ and $\overline{\alpha}_{t}:=\prod_{s=1}^{t}\alpha_{s}$. $\bm{\epsilon}_{\theta}^{(t)}$ is a function whose input and output sizes are the same as that of $\bm{\mu}_{\theta}$. $\bm{\Sigma}_{\theta}^{(t)}(\bm{x}^{(t)})$ is often set to be $\beta_{t}\bm{I}$ or $\tilde{\beta_{t}}\bm{I} = \frac{1-\overline{\alpha}_{t-1}}{1-\overline{\alpha}_{t}}\beta_{t}\bm{I}$. With this parameterization, the training objective is reduced to
\begin{align}
    \mathbb{E}_{\bm{x}^{(0)}, \bm{\epsilon}\sim\mathcal{N}(\bm{0}, \bm{I}), t} \left[\lVert\bm{\epsilon}-\bm{\epsilon}_{\theta}^{(t)}(\sqrt{\overline{\alpha}_{t}}\bm{x}^{(0)} + \sqrt{1-\overline{\alpha}_{t}}\bm{\epsilon})\rVert_{2}^{2} \right].
\end{align}
Note that with the parameterization of \eqref{eq:def_mu}, the function $f_{\theta}^{(t)}(\bm{x}^{(t)}) := (\bm{x}^{(t)}-\sqrt{1-\overline{\alpha}_{t}}\bm{\epsilon}_{\theta}^{(t)}(\bm{x}^{(t)}))/\sqrt{\overline{\alpha}_{t}}$ can be viewed as a denoiser at each time step $t$, which can predict the clean signals $\bm{x}^{(0)}$ given $\bm{x}^{(t)}$~\cite{ho2020denoising}.

\subsection{Denoising diffusion restoration models}
Kawar \textit{et al.} proposed DDRM~\cite{kawar2022denoising}, which is an unsupervised method for solving general linear inverse problems. 
The goal of a linear inverse problem is to restore the signal $\bm{x}\in\mathbb{R}^{n}$ from the observation $\bm{y}\in\mathbb{R}^{m}$ obtained by the following linear equation:
\begin{align}
    \bm{y} = \bm{H}\bm{x}+\bm{z},
    \label{eq:general_inverse_problems}
\end{align}
where $\bm{z}\sim \mathcal{N}(0, \sigma_{\bm{y}}^{2}\bm{I})$ is an \textit{i.i.d.} additive Gaussian noise with known variance.
Here, $\bm{H}\in\mathbb{R}^{m\times n}$ is a degradation linear operator and is assumed to be known.

For any linear inverse problem, the DDRM is defined as
\begin{align}
    p_{\theta}(\bm{x}^{(0:T)}|\bm{y}) = p_{\theta}^{(T)}(\bm{x}^{(T)}|\bm{y})\prod_{t=0}^{T-1}p_{\theta}^{(t)}(\bm{x}^{(t)}|\bm{x}^{(t+1)}, \bm{y}),
\end{align}
where $\bm{x}^{(0)}$ is the estimate of $\bm{x}$ in this model. Given the singular value decomposition (SVD) of $\bm{H}$, DDRM can take how much information from $\bm{y}$ is available into account in the domain induced by the SVD (for denoising).
For instance, components corresponding to larger singular values are less-noisy in the spectral space, and thus much information from the observation is used to restore the signal. 
Meanwhile, components corresponding to smaller singular values are hard to observe because of lower SNR, and thus much information from the generative model is used.


The point of DDRM is that once an unconditional diffusion-based generative model is trained on clean data, it can be utilized for various types of linear inverse problems because the knowledge of the linear operator is required only at an inference time, as discussed in the original paper~\cite{kawar2022denoising}.

\vspace{-1mm}
\section{Diffusion-based Speech Refiner}
\label{sec:propose}
\vspace{-0.75mm}
\begin{figure}[!t]
  \centering
    \includegraphics[width=8.0cm]{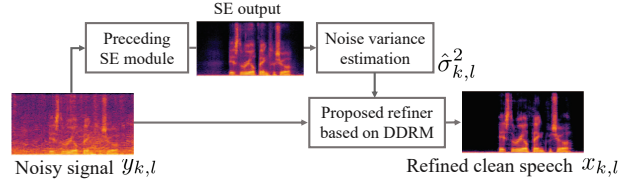}
    \vspace{-5mm}
    \caption{Overview of proposed diffusion-based speech refiner}
    \vspace{-4mm}
    \label{fig:overview}
\end{figure}
In this section, we propose our diffusion-based refiner for SE outputs. First, we train a diffusion-based generative model on clean speech data. After obtaining results from an \nm{arbitrary preceding SE module}, the variance of the noise included in noisy input at each time-frequency bin is estimated. With the estimate, the proposed refiner generates clean speech on the basis of the DDRM framework, which utilizes the pre-trained diffusion-based model.

\vspace{-1mm}
\subsection{Diffusion-based generative model on clean speech data}
\label{sec:3}
\vspace{-0.5mm}
We assume that the noisy time-domain signal $y$ is decomposed as $y = x + n$, where $x$ and $n$ are the target signals (to be enhanced) and noise signals, respectively. Then we can obtain their short-time Fourier transformation (STFT) coefficients, denoted as $y_{k, l}, x_{k, l}, \text{and }n_{k, l}\in\mathbb{C}$, respectively. $k$ and $l$ denote the frequency bin and time frame indexes.
Note that the existing diffusion-based models can be extended to complex-valued data, such as STFT coefficients~\cite{welker2022speech, richter2022speech}.

First, a diffusion-based generative model for the STFT coefficients is trained on clean speech data, which gives the pre-trained diffusion-based generative model $x_{k, l}\sim p_{\theta}(x_{k, l})$. Here, the forward process for the Markov chain is defined as the process that injects circular-symmetric complex Gaussian noise as follows,
\begin{align}
    q^{(t)}(x_{k, l}^{(t)}\mid x_{k, l}^{(0)}) = \mathcal{N}_{\mathbb{C}}(x_{k, l}^{(0)}, \sigma_{t}^{2}), 
\end{align}
with different noise levels $0=\sigma_{0}<\sigma_{1}<\dots<\sigma_{T}$. As in general diffusion-based generative models, although this forward process that adds noise is performed independently for each time-frequency bin, the reverse process \nm{that removes noise} is trained by considering correlations between the bins. Again, note that, with the trained model $p_{\theta}(x_{k, l})$, the denoiser $f_{\theta}^{(t)}(\cdot)$ that is defined in the same manner as Sec.~\ref{sec:DDPM} can predict the denoised signals, which is denoted here as $\overline{x}_{k, l}^{(t)}$, given noisy signals $x_{k, l}^{(t)}$.

\subsection{Diffusion-based refiner for SE outputs}
\label{subsec: refiner}
In this subsection, we introduce our DDRM-based refiner. After obtaining the clean speech estimation with the existing SE module, which we denote as $\hat{x}_{k, l}$, we can also compute the estimated noise signal $\hat{n}_{k, l}$ by subtracting $\hat{x}_{k, l}$ from $y_{k, l}$. Then we model the estimated variance of STFT coefficients $\hat{n}_{k, l}$, as
\begin{align}
    \hat{\sigma}_{k, l}^{2} = \min(\max (\lambda|\hat{n}_{k, l}|^{2}, \delta), R)
\end{align}
where $\delta > 0$ and $R > 0$ are the minimum and maximum thresholds of the estimated variance for avoiding numerical instability, respectively. We leave $\lambda > 0$ as a tunable hyperparameter.
This variance estimate leads to the following approximated linear inverse problem:
\begin{align}
    y_{k, l} = x_{k, l} + n_{k, l}, \ n_{k, l}\sim \mathcal{N}_{\mathbb{C}}(0, \hat{\sigma}_{k, l}^{2}), 
\end{align}
and it can be written in the form of a general linear inverse problem as in~\eqref{eq:general_inverse_problems}:
\begin{align}
    \tilde{y}_{k, l} = \frac{1}{\hat{\sigma}_{k, l}}x_{k, l} + z_{k, l}, z_{k, l} \sim \mathcal{N}_{\mathbb{C}}(0, 1), 
    \label{eq:res_inv_prob}
\end{align}
where $\tilde{y}_{k, l} = y_{k, l}/\hat{\sigma}_{k, l}$. Here, we assume $z_{k, l}$ follows an \textit{i.i.d.} complex Gaussian distribution for tractability. Based on the inverse problem in~\eqref{eq:res_inv_prob}, the proposed method is able to generate refined signals with DDRM, which is summarized in Algorithm~\ref{alg:proposal}. \nm{$\eta_{a}, \eta_{b},$ and $\eta_{c} \in [0, 1]$ are tunable hyperparameters that control diversity of generated samples.} For notational simplicity, $\tilde{y}_{k, l}$ is replaced by $y_{k, l}/\hat{\sigma}_{k, l}$ in the algorithm. We can interpret this to mean that the algorithm generates speech by appropriately combining the estimated clean speech $\overline{x}_{k, l}^{(t)}$ with the observed noisy speech $y_{k, l}$ in accordance with the estimated noise variance $\hat{\sigma}_{k, l}^{2}$ at each time-frequency bin.

In practical cases of SE, the estimated variance $\hat{\sigma}_{k, l}^2$ may include estimation errors and additive noise $z_{k, l}$ may have correlations among different time-frequency bins. Thus, the term $(y_{k, l}-\overline{x}_{k, l}^{(t)})$ in the update procedure, which uses the information from the observation, does not necessarily follow the independent complex Gaussian with the estimated variance, which is assumed in DDRM. To address this issue, we add a modification to the original DDRM framework
as shown in Algorithm~\ref{alg:proposal}, which is referred to as \textbf{``Diffiner+"}. We use the term $(x_{k, l}^{(t+1)}-\overline{x}_{k, l}^{(t)})$ instead of $(y_{k, l}-\overline{x}_{k, l}^{(t)})$ if $\sigma_{t} < \hat{\sigma}_{k, l}$ because the observed signal is no longer helpful in this condition. This change in the term means that the algorithm ignores the observational information and relies on the generative model during less noisy steps ($\sigma_{t} < \hat{\sigma}_{k, l}$).

\begin{algorithm}
\caption{Proposed diffusion-based refiner for SE outputs}\label{alg:proposal}
\begin{algorithmic}
\Require noisy input $y_{k, l}$, estimated noise variance $\hat{\sigma}_{k, l}^2$
\Ensure refined signal $x_{k, l}$
\State initialize $x_{k, l}^{(T)}\sim \mathcal{N}_{\mathbb{C}}(0, \sigma_{T}^2- \hat{\sigma}_{k, l}^{2})$
\For {$t=T-1$ to $0$}
    \State Predict denoised signal $\overline{x}_{k, l}^{(t)}$ using $f_{\theta}^{(t)}(\cdot)$
    \State Sample $z_{k, l}\sim\mathcal{N}_{\mathbb{C}}(0, 1)$
    \If{$\sigma_{t}<\hat{\sigma}_{k, l}$}
        \State \underline{Pattern 1: Original DDRM update rule (\textbf{Diffiner})}
        \State $x_{k, l}^{(t)} = \overline{x}_{k, l}^{(t)}+\eta_{a}\sigma_{t}\frac{y_{k, l}-\overline{x}_{k, l}^{(t)}}{\hat{\sigma}_{k,l}}+\sqrt{1-\eta_{a}^2}\sigma_{t}z_{k, l}$
        \State \underline{Pattern 2: Modified update rule (\textbf{Diffiner+})} 
        \State $x_{k, l}^{(t)} = \overline{x}_{k, l}^{(t)}+\eta_{c}\sigma_{t}\frac{x_{k, l}^{(t+1)}-\overline{x}_{k, l}^{(t)}}{\sigma_{t+1}}+\sqrt{1-\eta_{c}^2}\sigma_{t}z_{k, l}$
    \Else \ \ \texttt{//} $\sigma_{t}\geq\hat{\sigma}_{k, l}$
        \State $x_{k, l}^{(t)} = (1-\eta_{b})\overline{x}_{k, l}^{(t)}+\eta_{b}y_{k, l}+\sqrt{\sigma_{t}^2-\eta_{b}^{2}\hat{\sigma}_{k, l}^{2}}z_{k, l}$
    \EndIf
\EndFor
\State Refined signals $x_{k, l} = x_{k, l}^{(0)}$
\end{algorithmic}
\end{algorithm}

\section{Experiments}
\label{sec:exp}

\subsection{Setup}
\label{subsec:exp_setting}
\subsubsection{Dataset}
\label{subsubsec:exp_data}
To train and evaluate our proposed model, we utilized an openly available dataset, Voice Bank Corpus (VBC) \cite{voicebank}, consisting of only clean speech since our diffusion-based refiner does not require noisy or noise signals as we discussed in Sec.~\ref{sec:intro}.

Meanwhile, to train and evaluate DNN-based SE models that are used before our refiner, we utilized VoiceBank-DEMAND (VBD) \cite{voicebankdemand}, which is also openly available and frequently used in DNN-based speech enhancement \cite{dcunet_1,segan}.
The train and test sets consist of 28 and two speakers (11572 and 824 utterances), respectively.

\subsubsection{Preceding SE methods}
\label{subsubsec:preproc}
We used four SE methods as preceding modules to evaluate our refiner: Wiener filter \cite{wiener_org}, Deep complex U-net (DCUnet) \cite{dcunet_1}, Wave-U-net \cite{waveunet}, and speech enhancement GAN (SEGAN) \cite{segan}.
Note that all of them except the Wiener filter were trained on the VBD dataset.

\subsubsection{Proposed diffusion-based generative refiner}
\label{subsubsec:diffrefiner}
For the proposed refiner, we trained a diffusion-based generative model on STFT spectrograms of the clean speech obtained from the VoiceBank corpus.
As parameters for STFT, the window size, hop size, and number of time frames were set to 512, 256, and 256, respectively. 
A Hann window was used as the analysis window. By truncating the direct current (DC) component, we treated the spectrograms as $256\times 256$ tensors with two channels, which correspond to the real part and the imaginary part of a complex value. 
We modified the diffusion-based model with U-Net-based architecture\footnote{\url{https://github.com/openai/guided-diffusion}} so that the number of input/output channels is two.

We trained the model on a single NVIDIA A100 GPU (40~GB memory) for $7.5\times10^{5}$ steps, which took about three days. We used the Adam optimizer~\cite{adam} with a learning rate of 0.001 and batch size of 8. Following \cite{song2020improved}, an exponential moving average of the model weight was taken with a decay of 0.9999 to be used for inference.

For the inference, we first conducted a grid search for the parameters required in both of the deep generative ``Diffiner'' and ``Diffiner+'': $\eta_a$, $\eta_b$, and $\eta_c$ in Algorithm~\ref{alg:proposal}.
Specifically, we searched the parameter space $[0.0, 0.2, 0.4, 0.6, 0.8,$ $1.0]$ for $\eta_a$, $\eta_b$, and $\eta_c$, respectively.
Furthermore, $\lambda$, $\delta$, and $R$ were set to $\lambda=1.0$, $\delta=1.0^{-5}$, and $R=\sigma_{T-1}^{2}\simeq 97$, respectively.
Then, we ran our refiners with $T=200$. 

\subsection{Results}
\label{exp_results}
%
\begin{table}[!t]
\centering
\caption{
    Experimental results.
    Please note that the details of our proposed Diffiner and Diffiner+ are written in Sec.~\ref{subsec: refiner} and Algorithm~\ref{alg:proposal}.
}
\vspace{-3.0mm}
\resizebox{8.0cm}{!}{
\begin{threeparttable}
\begin{tabular}{ c | c c | c c c }
	\hline
    \multirow{2}{*}{\textbf{Method}}
    & \multicolumn{2}{c|}{\textbf{Reference-free Metrics}} 
    & \multicolumn{3}{c}{\textbf{Reference-based Metrics}} \\ \cline{2-6}
    %
         & \textbf{NISQA} & \textbf{OVRL}
         & \textbf{SI-SDR} & \textbf{WB-PESQ} &\textbf{ESTOI} \\ \hline \hline
Source & 4.546 & 3.220 & - & 4.644 & 1.000 \\
Input (noisy) & 3.040 & 2.697 & 8.448 & 1.971 & 0.787 \\ \hline
\multicolumn{1}{l|}{Wiener filter \cite{wiener_org}} & 3.544 & 2.846 & 15.65 & 2.414 & 0.793 \\
\multicolumn{1}{l|}{\hspace{+2.1mm}w/ Diffiner} & 4.472 & 3.064 & \textbf{18.09} & \textbf{2.476} & \textbf{0.847} \\
\multicolumn{1}{l|}{\hspace{+1.2mm} w/ Diffiner+} & \textbf{4.621} & \textbf{3.079} & 16.52 & 2.387 & 0.820 \\ \hdashline
\multicolumn{1}{l|}{DCUnet \cite{dcunet_1}} & 4.287 & 3.149 & \textbf{20.16} & \textbf{2.981} & \textbf{0.886} \\
\multicolumn{1}{l|}{\hspace{+2.1mm}w/ Diffiner} & 4.752 & 3.183 & 19.80 & 2.970 & 0.885 \\
\multicolumn{1}{l|}{\hspace{+1.2mm} w/ Diffiner+} & \textbf{4.827} & \textbf{3.187} & 19.27 & 2.810 & 0.861 \\ \hdashline
\multicolumn{1}{l|}{Wave-U-net \cite{waveunet}} & 3.968 & 3.091 & 18.15 & 2.657 & 0.842 \\
\multicolumn{1}{l|}{\hspace{+2.1mm}w/ Diffiner} & 4.663 & 3.159 & \textbf{19.62} & \textbf{2.684} & \textbf{0.871} \\
\multicolumn{1}{l|}{\hspace{+1.2mm} w/ Diffiner+} & \textbf{4.773} & \textbf{3.161} & 18.33 & 2.608 & 0.848 \\ \hdashline
\multicolumn{1}{l|}{SEGAN \cite{segan}} & 3.527 & 3.019 & 15.94 & 2.166 & 0.823 \\
\multicolumn{1}{l|}{\hspace{+2.1mm}w/ Diffiner} & 4.372 & 3.154 & \textbf{19.36} & \textbf{2.546} & \textbf{0.867} \\
\multicolumn{1}{l|}{\hspace{+1.2mm} w/ Diffiner+} & \textbf{4.609} & \textbf{3.160} & 17.79 & 2.513 & 0.851 \\ \hline
UNIVERSE\tnote{$\dag$}\hspace{+1.5mm}\cite{universe_se_arxiv} & 4.606 & 3.109 & 10.10 & 2.901 & 0.838 \\ 
SGMSE+\tnote{$\ddag$}\hspace{+1.5mm}\cite{richter2022speech} & 4.565 & 3.178 & 17.42 & 2.903 & 0.864 \\ 
\hline
\end{tabular}
\begin{tablenotes}
\fontsize{10pt}{0cm}\selectfont
    \item[$\dag$]
        Implemented by ourselves because the code is not publicly available. The network was trained on VBD, and we only considered Mel bands for feature NLLs.
    \item[$\ddag$]
        We used the authors' code and a provided checkpoint, but some results were slightly different from their paper's evaluation scores.
        This is because the predictor-corrector samplers used in SGMSE+ are stochastic.
\end{tablenotes}
\end{threeparttable}
\label{tb:exp_results}
}
\vspace{-2.0mm}
\end{table}
%
To evaluate the performance, we used 5 metrics\footnote{To compute these metrics, we utilized the latest versions of the following public codes as of March 2023:\\
$\cdot$\url{https://github.com/sigsep/sigsep-mus-eval}\\
$\cdot$\url{https://github.com/gabrielmittag/NISQA}\\
$\cdot$\url{https://github.com/microsoft/DNS-Challenge}}: SI-SDR, WB-PESQ, ESTOI, non-intrusive speech quality assessment (NISQA), and the overall (OVRL) metric of the deep noise challenge mean opinion score (DNSMOS) P.835 \cite{sisdr, pesq, estoi, nisqa_org, dnsmos_p835}.
The first 3 metrics, which require the pair of a target signal and the corresponding reference source, have been recently used for performance evaluation on VBD \cite{richter2022speech, diffsep}.
However, we focus on improving the last 2 metrics, i.e., NISQA and OVRL, in our experiments.
This is because some methods for speech enhancement, including our refiner, are based on generative models, whose outputs contain generated parts that do not completely match the corresponding references.
In contrast with the aforementioned traditional reference-based metrics, NISQA and OVRL can predict the mean opinion score (MOS) of a target signal without the corresponding reference source.
Thus, we considered that they are suitable for the evaluation of the generative model-based results.

The evaluation results are summarized in Table~\ref{tb:exp_results}.
Note that we added some diffusion-based state-of-the-art methods, universal speech enhancement with score-based diffusion (UNIVERSE) \cite{universe_se_arxiv} and the improved version of score-based generative model for SE (SGMSE+) \cite{richter2022speech}, as references.
As shown in the table, all reference-free scores, i.e., NISQA and OVRL, were always improved by applying our refiners, effectively.
Therefore, in terms of human-like MOS scores, both Diffiner and Diffiner+ succeeded to improve the SE results regardless of what kind of preceding SE method was used.
In particular, the results obtained by applying Diffiner+ are comparable to the source in NISQA and OVRL.
Namely, in terms of reference-free metrics related to MOS, Diffiner+ could improve all preceding SE methods, achieving as natural speech as the source.
Furthermore, all NISQA and OVRL scores by Diffiner+ are comparable to or higher than those of UNIVERSE and SGMSE+, which are state-of-the-art diffusion-based SE methods.
In contrast, our refiners reduced the reference-based metrics marginally.
As we discussed in the previous paragraph, this is because the refined results contain generated parts that do not match the reference, and thus it could spoil the reference-based scores.
However, as shown in the table, Diffiner+ could keep its scores comparable with that of the source while improving NISQA and OVRL.
Thus, although it is a generative model, Diffiner+ can refine the target signals and keep them close to the corresponding references.
\begin{figure}[!t]
  \centering
  \subfloat[Input (noisy)]{
    \includegraphics[width=3.9cm, height=1.7cm]{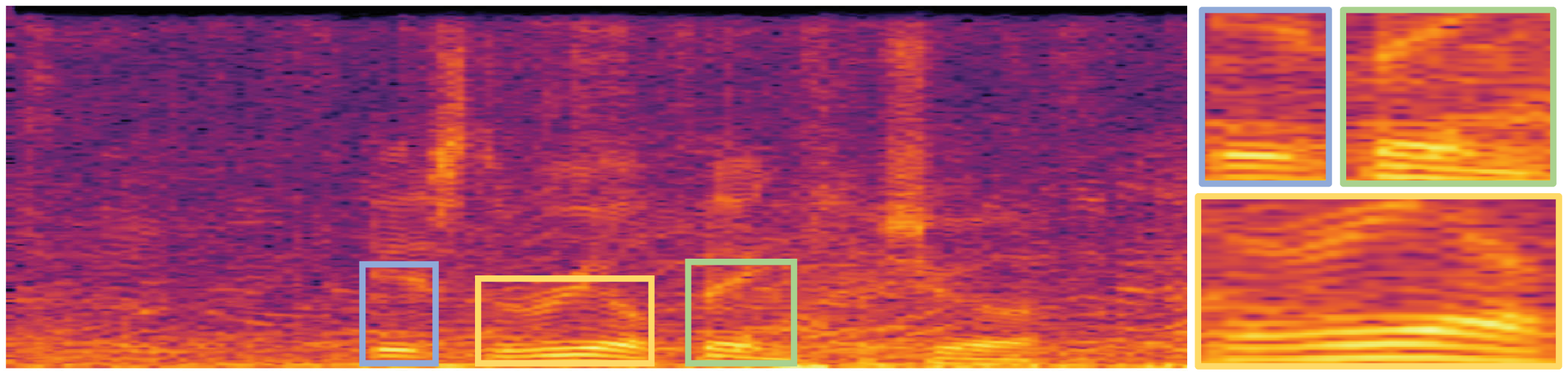}    \label{fig:input_spec}}
  \subfloat[Source]{
    \includegraphics[width=3.9cm, height=1.7cm]{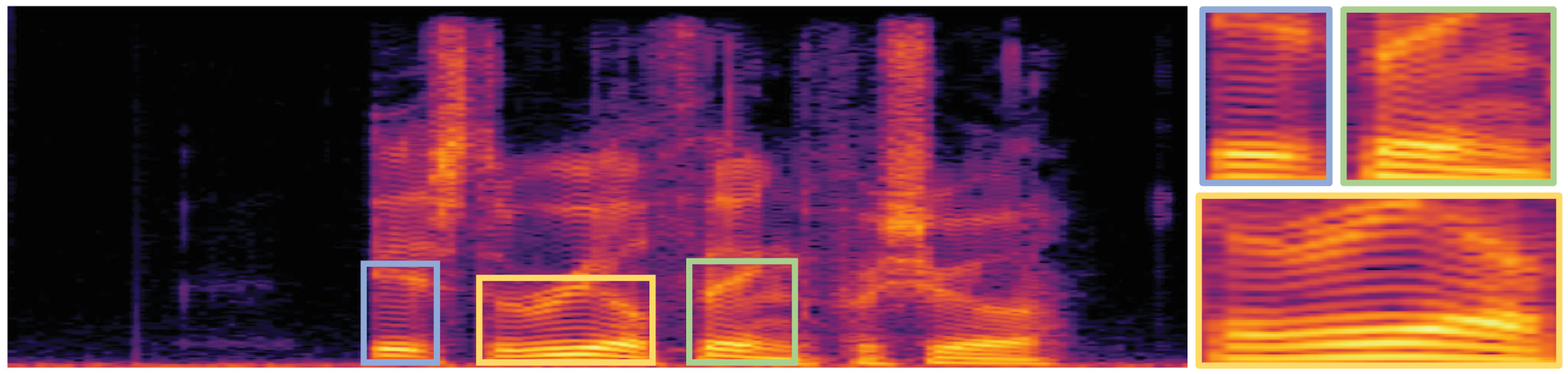}
    \label{fig:clean_spec}}

  \vspace{-2mm}
  \subfloat[Pre-processed by DCUnet]{
    \includegraphics[width=3.9cm, height=1.7cm]{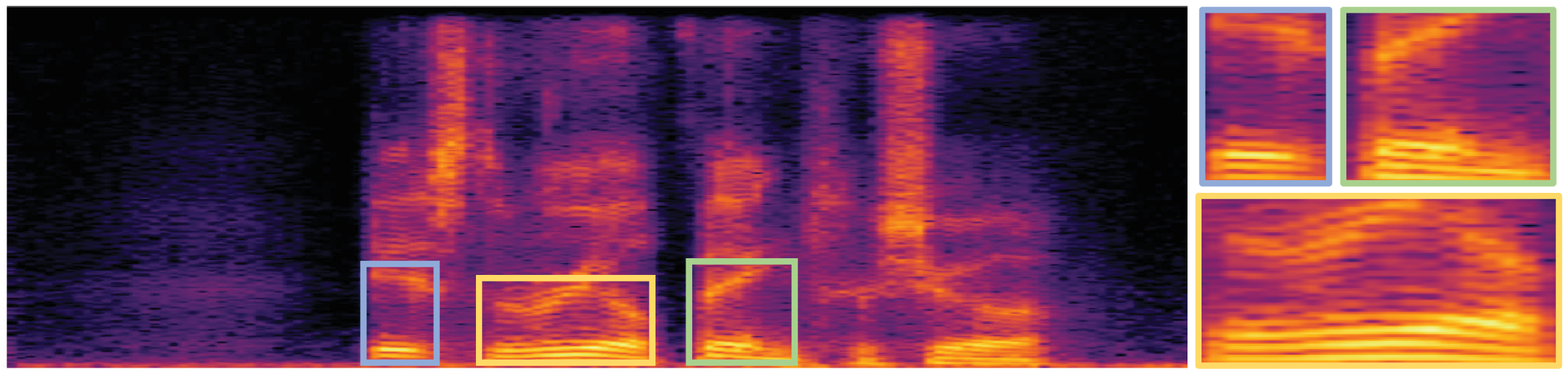}
    \label{fig:dcunet_spec}}
  \subfloat[Refined by Diffiner+]{
    \includegraphics[width=3.9cm, height=1.7cm]{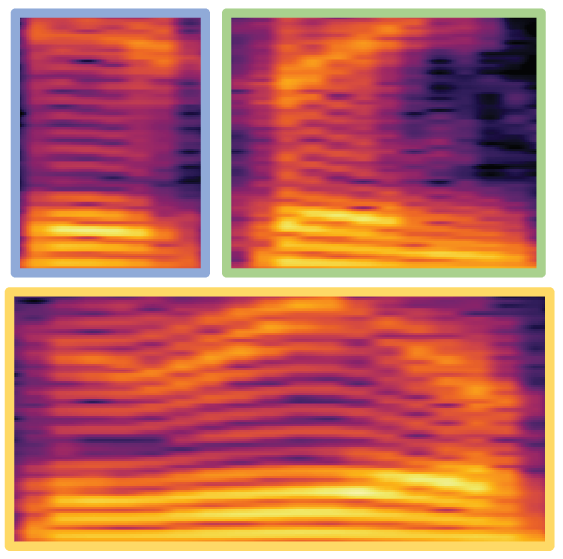}
    \label{fig:proposed_spec}}
  \vspace{-2.5mm}
  \caption{Spectrograms of noisy input, clean target, DCUnet, and refined by diffusion-based generative refiner.}
  \vspace{-2.5mm}
  \label{fig:comp_spec}
\end{figure}

An example of the refined results is shown in Fig.~\ref{fig:comp_spec}. 
Even though the details of the refined parts were not always consistent with the corresponding parts of the source spectrogram (see Figs.~\ref{fig:comp_spec}(b) and (d)), our refiner could generate natural-looking parts and
succeeded to mix them into the distorted parts resulting in a high-quality spectrogram (see Figs.~\ref{fig:comp_spec}(c) and (d)).

\section{Conclusion}
\label{sec:conc}
We presented a DNN-based generative refiner to improve speech that has already been pre-processed by a preceding SE method.
We devised ``Diffiner'', an extension of DDRM, that is more suitable for the task of speech enhancement.
After our model is trained on a set of clean speech, it can be used as a versatile speech refiner for results processed by preceding SE methods without additional training specialized for each method.
Experimental results showed that our method effectively improved the speech quality in terms of NISQA and OVRL regardless of the SE method used as pre-processing.
In future work, we will explore the feasibility of our deep generative refiner for a scaled dataset larger than VBC and its application to other types of sound such as music and environmental sounds.
Moreover, our refiner can be integrated with various DNN-based SE methods, resulting in a joint generative refiner.
Especially, there may be an effective way to integrate our deep generative refiner with other diffusion-based SE methods \cite{se_diffusion_1, se_diffusion_2, se_diffusion_3, universe_se_arxiv, richter2022speech} into a single unified model.

\clearpage
\bibliographystyle{IEEEtran}
\bibliography{ref, ref_dgm}
\end{document}